%% file: main.tex
\documentclass[conference]{IEEEtran}
\IEEEoverridecommandlockouts
\usepackage{cite}
\usepackage{amsmath,amssymb,amsfonts}
\usepackage{algorithmic}
\usepackage{graphicx}
\usepackage{textcomp}
\usepackage{xcolor}
\usepackage{booktabs}
\usepackage{multirow}
\usepackage{graphicx}

\begin{document}

\title{Cross-Domain Image Synthesis: Generating H\&E from Multiplex Biomarker Imaging}

\input{texes/authors}



\maketitle

\input{texes/0_abstract}


\input{texes/1_intro}
\input{texes/2_background_study}
\input{texes/3_models}
\input{texes/5_results}
\input{texes/6_conclusion}
\bibliographystyle{IEEEtran}
\bibliography{mybib}

\end{document}

%% file: texes/authors.tex
\author{
    \IEEEauthorblockN{Jillur Rahman Saurav}
    \IEEEauthorblockA{Department of CSE\\
    University of Texas at Arlington\\
    Arlington, Texas, USA\\
    Email: mxs2361@mavs.uta.edu}
    \and
    \IEEEauthorblockN{Mohammad Sadegh Nasr}
    \IEEEauthorblockA{Department of CSE\\
    University of Texas at Arlington\\
    Arlington, Texas, USA\\
    Email: mohammadsadegh.nasr@mavs.uta.edu}
    \and
    \IEEEauthorblockN{Jacob M. Luber}
    \IEEEauthorblockA{Department of CSE\\
    University of Texas at Arlington\\
    Arlington, Texas, USA\\
    Email: jacob.luber@uta.edu}
}

%% file: texes/0_abstract.tex
\begin{abstract}
While multiplex immunofluorescence (mIF) imaging provides deep, spatially-resolved molecular data, integrating this information with the morphological standard of Hematoxylin \& Eosin (H\&E) can be very important for obtaining complementary information about the underlying tissue. Generating a virtual H\&E stain from mIF data offers a powerful solution, providing immediate morphological context. Crucially, this approach enables the application of the vast ecosystem of H\&E-based computer-aided diagnosis (CAD) tools to analyze rich molecular data, bridging the gap between molecular and morphological analysis. In this work, we investigate the use of a multi-level Vector-Quantized Generative Adversarial Network (VQGAN) to create high-fidelity virtual H\&E stains from mIF images. We rigorously evaluated our VQGAN against a standard conditional GAN (cGAN) baseline on two publicly available colorectal cancer datasets, assessing performance on both image similarity and functional utility for downstream analysis. Our results show that while both architectures produce visually plausible images, the virtual stains generated by our VQGAN provide a more effective substrate for computer-aided diagnosis. Specifically, downstream nuclei segmentation and semantic preservation in tissue classification tasks performed on VQGAN-generated images demonstrate superior performance and agreement with ground-truth analysis compared to those from the cGAN. This work establishes that a multi-level VQGAN is a robust and superior architecture for generating scientifically useful virtual stains, offering a viable pathway to integrate the rich molecular data of mIF into established and powerful H\&E-based analytical workflows.
\end{abstract}

%% file: texes/1_intro.tex
\section{Introduction}

Modern pathology stands at the convergence of two powerful imaging paradigms: morphological assessment and high-dimensional molecular mapping \cite{lin2023high}. Hematoxylin and Eosin (H\&E) staining has served as the foundation for cancer diagnosis for over a century, enabling the visual assessment of cellular and tissue architecture.(H\&E)\cite{gurcan2009histopathological}. This foundational technique provides the morphological context that underpins routine diagnostic pathology and has given rise to a vast ecosystem of validated analytical tools \cite{chan2014wonderful, madabhushi2016image}. In parallel, technologies such as multiplex immunofluorescence (mIF) have enabled the era of spatial biology, offering detailed views of the tumor microenvironment by visualizing dozens of protein biomarkers within a single tissue sample \cite{giesen2014highly}.
\begin{figure}[t]
    \centering
    \includegraphics[width=\columnwidth]{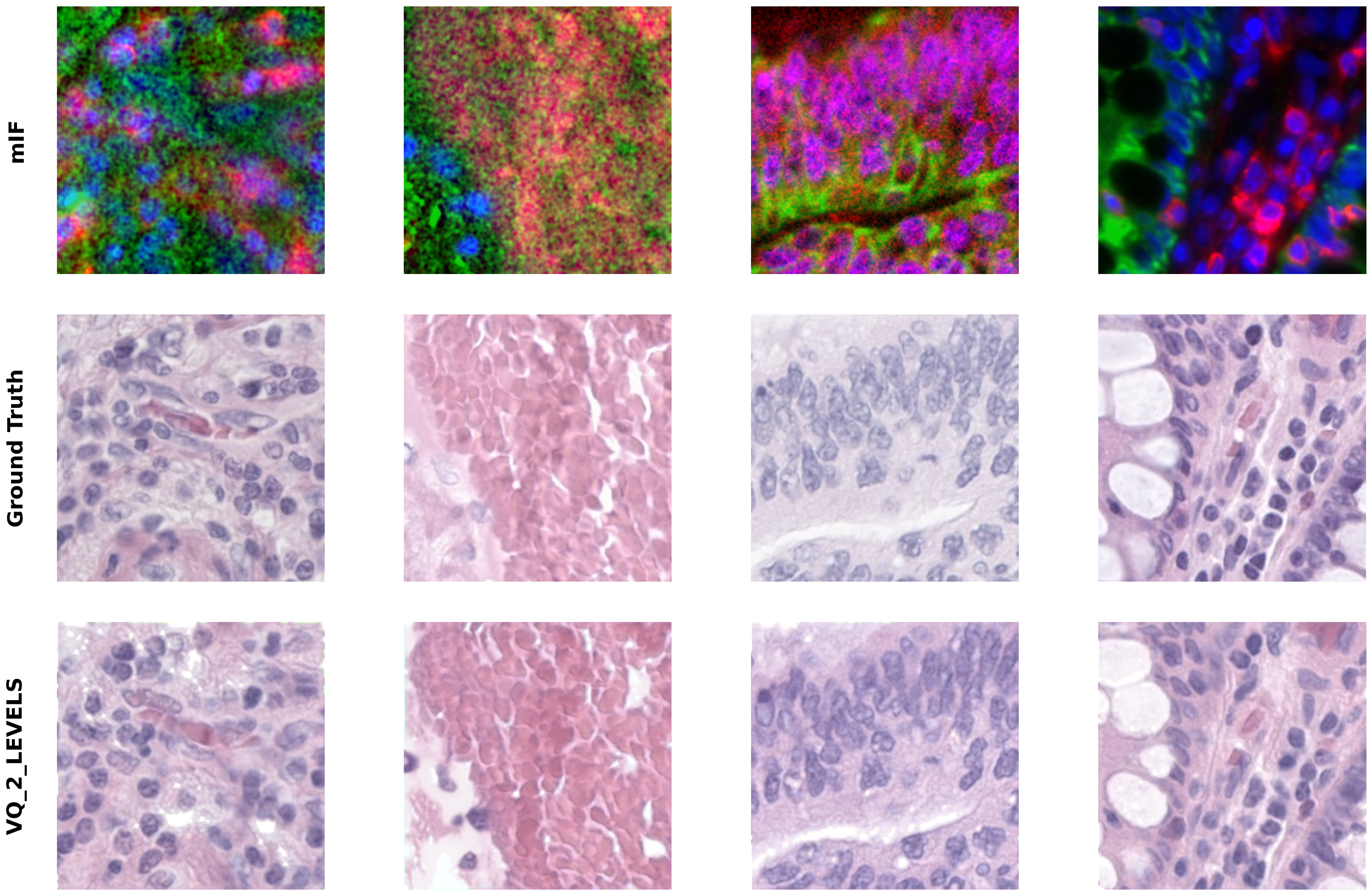}
    \caption{
        \textbf{High-Fidelity Virtual H\&E Staining using a Multi-Level VQGAN.} Representative examples from the Orion colorectal cancer dataset \cite{lin2023high}. Each row displays: (a) the input multiplex immunofluorescence (mIF) image (R: CD31, G: E-cadherin, B: Hoechst), (b) the ground truth H\&E stain, and (c) the H\&E image generated by our proposed model.
    }
    \label{fig:overview}
\end{figure}
While mIF provides unprecedented molecular depth, its high-dimensional data format is incompatible with the extensive computational infrastructure built for H\&E analysis. Moreover, each modality reveals complementary information: mIF excels at identifying immune cell subtypes and protein expression patterns that are morphologically indistinguishable in H\&E, while H\&E provides superior visualization of tissue architecture, acellular structures, and morphological features that may lack specific molecular markers \cite{lin2023high}. This disconnect limits the ability to fully leverage the complementary information the rich molecular data from mIF cannot readily leverage the robust morphological analysis pipelines developed for H\&E, and pathologists cannot easily apply their H\&E-based expertise to interpret mIF imagery directly.

To bridge this gap, computational pathology has increasingly focused on cross-domain image synthesis, or "virtual staining" \cite{rivenson2019virtual}. Generative Adversarial Networks (GANs) have been the dominant approach for this task. For unpaired data scenarios, such as translating between H\&E and single-channel immunohistochemistry (IHC) stains, methods like CycleGAN have been employed \cite{bao2024alleviating}. For paired data settings, conditional GANs (cGANs) have demonstrated strong performance in various stain-to-stain translation tasks \cite{saad2025automatic}. Despite their success, GAN-based approaches face inherent challenges, including training instability and potential mode collapse \cite{cobbinah2025diversity}.

An alternative paradigm, vector quantization, offers a potentially more stable approach to image synthesis. Vector Quantized Variational Autoencoders (VQ-VAEs) and their GAN-based extensions (VQGANs) learn discrete, compositional representations of data through learned codebooks \cite{oord2017neural,esser2021taming}. This discretization can lead to more stable training and encourages the model to capture recurring visual patterns as distinct codebook entries. Hierarchical VQ architectures, in particular, have shown success in modeling information at multiple spatial scales in natural images \cite{razavi2019generating}, a concept that aligns well with how pathologists analyze tissue at varying levels of magnification—from tissue architecture to cellular morphology to subcellular details.

Despite these theoretical advantages, vector quantization has not been systematically evaluated for medical image translation tasks, particularly for the challenging mIF-to-H\&E synthesis problem. The high-dimensional nature of mIF data (often 19-58 channels) coupled with the need to preserve precise spatial relationships between molecular markers and morphological features presents unique technical challenges that require specialized architectural innovations.

In this paper, we investigate the potential of this discrete-latent paradigm for the challenging task of mIF-to-H\&E virtual staining. Our contributions are threefold: First, to our knowledge, this work represents the first systematic comparison of generative models for direct, paired mIF-to-H\&E translation. Second, we demonstrate through a rigorous head-to-head comparison that a multi-level VQGAN yields images that are more effective for downstream scientific analysis than a standard cGAN, achieving superior performance on segmentation and semantic preservation tasks. Finally, by validating the functional utility of the generated stains through downstream analysis, we establish the VQGAN as a promising and effective architecture for integrating rich multiplex imaging data into established pathology workflows.

%% file: texes/2_background_study.tex
\section{Related Work}

\subsection{Virtual Staining and Cross-Modal Translation}

Virtual staining has emerged as a transformative approach to bridge the gap between different imaging modalities in pathology. Early pioneering work by Rivenson et al. \cite{rivenson2019virtual} demonstrated that deep learning could transform autofluorescence images of unlabeled tissue into histologically stained equivalents, with pathologist validation showing no major discordances compared to traditional staining. Christiansen et al. \cite{christiansen2018silico} introduced the concept of "in silico labeling," showing that machine learning could predict fluorescent labels from transmitted-light images without physical labeling, addressing key limitations such as spectral overlap and experimental perturbation.

The field has since expanded to various stain-to-stain transformations, particularly H\&E to immunohistochemical (IHC) translation. Recent work by Chen et al. \cite{chen2024pathological} proposed pathological semantics-preserving learning to address spatial misalignment issues in H\&E-to-IHC translation, while Kataria et al. \cite{kataria2024staindiffuser} introduced diffusion-based approaches for virtual IHC generation.Bao et al. \cite{bao2024alleviating} represents the only prior work we identified that addresses virtual H\&E generation from multiplex immunofluorescence data. They proposed an unpaired high-resolution image synthesis method to obtain virtual H\&E whole slide images from multiplex immunofluorescence images with 27 markers, focusing on reducing tiling effects through spatial constraints and sliding window strategies. However, their approach utilized unpaired data and CycleGAN-based training, making direct comparison with our paired, supervised approach unsuitable.

Despite these advances, most virtual staining work has focused on single-channel to single-channel transformations or relatively low-dimensional inputs. The translation from high-dimensional multiplex immunofluorescence (19-58 channels) to H\&E represents a significantly more challenging problem that requires specialized architectural considerations.

\subsection{Generative Models for Medical Image Synthesis}

Deep generative models have become the predominant approach for virtual staining applications. Conditional Generative Adversarial Networks (cGANs), based on the pix2pix framework \cite{isola2017image}, have been widely adopted for paired image-to-image translation tasks in medical imaging. The adversarial training paradigm enables the generation of realistic textures and fine-grained details essential for pathological analysis.

However, GAN-based approaches face inherent challenges in medical applications, including training instability, mode collapse, and sensitivity to hyperparameter selection \cite{saad2024survey}. Recent comprehensive reviews \cite{kazeminia2020gans} have highlighted performance variations across different datasets and the need for more robust architectural approaches. Additionally, the adversarial training process can sometimes introduce artifacts that may compromise downstream analytical tasks, particularly in quantitative applications requiring precise cellular morphology.

Vector quantization offers an alternative paradigm that addresses some of these limitations. Vector Quantized Generative Adversarial Networks (VQGANs) \cite{esser2021taming} combine the benefits of discrete latent representations with adversarial training, potentially offering more stable training dynamics and better preservation of fine-grained structures. The discrete codebook mechanism encourages the model to learn compositional representations, which may be particularly advantageous for capturing the recurring histological patterns present in tissue samples. However, vector quantization approaches have not been systematically evaluated for medical image translation tasks, particularly for the challenging high-dimensional mIF-to-H\&E synthesis problem.

%% file: texes/3_models.tex
\section{Methods}

\subsection{Datasets}

We evaluated our methods on two distinct colorectal cancer (CRC) datasets, each providing paired multiplex immunofluorescence (mIF) and hematoxylin and eosin (H\&E) stained images but featuring different data curation strategies and mIF channel depths. The Orion dataset \cite{lin2023high} and CODEX dataset \cite{schurch2020coordinated} were used for evaluation. All images were prepared as 224×224 pixel patches for training and evaluation. To ensure a rigorous and unbiased evaluation, both datasets employed a strict patient-level splitting protocol to prevent data leakage, with details summarized in Table~\ref{tab:dataset_splits}.

\textbf{Orion CRC Dataset:} This dataset, featuring 19-channel mIF data, was curated to optimize for stable model training through class-balancing. We employed a vision-language model, CONCH \cite{lu2024visual}, for zero-shot classification of H\&E patches into nine distinct tissue categories. The dataset was then sampled to create a balanced distribution of these tissue types.

\textbf{CODEX CRC Dataset:} This dataset provides deeper 58-channel mIF data and was curated to reflect a natural class distribution without artificial balancing. This presents a more realistic, real-world scenario where certain tissue types may be naturally over- or under-represented.

\begin{table}[h!]
\centering
\caption{Dataset characteristics and patient-level splits.}
\label{tab:dataset_splits}
\begin{tabular}{llcc}
\toprule
\textbf{Dataset} & \textbf{Split} & \textbf{\# Samples} & \textbf{mIF Channels} \\
\midrule
\multirow{3}{*}{Orion} & Train & 47,824 & \multirow{3}{*}{19} \\
 & Validation & 10,800 & \\
 & Test & 10,800 & \\
\midrule
\multirow{3}{*}{CODEX} & Train & 4,368 & \multirow{3}{*}{58} \\
 & Validation & 912 & \\
 & Test & 1,056 & \\
\bottomrule
\end{tabular}
\end{table}

\subsection{Model Selection}
To find an optimal method for mIF-to-H\&E virtual staining, we compared three generative models. We implemented a conditional Generative Adversarial Network (cGAN) as a strong supervised baseline. We then developed and compared VQGAN models based on vector quantization to evaluate the efficacy of discrete latent representations: a standard single-level VQGAN and a novel multi-level VQGAN. This selection allows for a direct comparison between a standard continuous latent space approach (cGAN) and discrete, hierarchical latent space approaches (VQGAN).

\subsection{Model Architectures}

\subsubsection{Conditional GAN}
Following the pix2pix framework \cite{isola2017image}, our cGAN uses a U-Net \cite{ronneberger2015u} based generator with skip connections to map the multi-channel mIF input directly to a 3-channel H\&E output. A PatchGAN discriminator operates on 70×70 image patches, providing a loss signal that enforces local textural realism.

\subsubsection{Single-Level VQGAN}
Our baseline VQGAN implementation follows the architecture from Esser et al. \cite{esser2021taming}. A convolutional encoder maps the mIF input to a latent feature map, which is then quantized element-wise using a single learned codebook of size 1024. A corresponding decoder reconstructs the H\&E image from these discrete latent codes. This model is trained with a composite loss function combining reconstruction, perceptual, style, and stain-specific losses, along with a small-weight adversarial loss that is ramped up during training to avoid hallucination artifacts.

\subsubsection{Multi-Level VQGAN}
Our proposed multi-level VQGAN extends the single-level baseline with hierarchical quantization. The encoder produces feature maps at two different spatial resolutions, and each is quantized by a separate, dedicated codebook of size 1024. This hierarchical approach is designed to allow the model to capture coarse, high-level tissue structures in the low-resolution latent space and fine-grained cellular details in the high-resolution latent space. Similar to the single-level baseline, this model uses the same composite loss with a small-weight adversarial component that is gradually ramped up during training. The decoder integrates information from all levels to reconstruct the final H\&E image.

\subsection{Loss Functions}

\subsubsection{Composite Loss Formulation}
Both VQGAN models were trained by optimizing a composite loss function, $\mathcal{L}_{\text{total}}$, designed to enforce structural accuracy, textural realism, and H\&E-specific color fidelity. The total objective is a weighted sum of five distinct loss terms:

\[
\mathcal{L}_{\text{total}} = \lambda_{\text{recon}}\mathcal{L}_{\text{recon}} + \lambda_{\text{perc}}\mathcal{L}_{\text{perc}} + \lambda_{\text{style}}\mathcal{L}_{\text{style}} + \lambda_{\text{stain}}\mathcal{L}_{\text{stain}} + \lambda_{\text{adv}}\mathcal{L}_{\text{adv}}
\]

where $\mathcal{L}_{\text{recon}}$ is an L1 reconstruction loss, $\mathcal{L}_{\text{perc}}$ is a perceptual loss using KimiaNet features \cite{riasatian2021fine}, $\mathcal{L}_{\text{style}}$ is a style loss using VGG19 Gram matrices \cite{gatys2015neural}, $\mathcal{L}_{\text{stain}}$ is a stain-specific loss based on color deconvolution inspired from \cite{macenko2009method}, and $\mathcal{L}_{\text{adv}}$ is an adversarial loss with small weight $\lambda_{\text{adv}}$ that is gradually ramped up during training to avoid hallucination artifacts:

\begin{align*}
\mathcal{L}_{\text{recon}} &= \mathbb{E}_{\mathbf{x},\mathbf{y}} \|\mathbf{y} - G(\mathbf{x})\|_1 \\
\mathcal{L}_{\text{perc}} &= \mathbb{E}_{\mathbf{x},\mathbf{y}} \sum_{i} \|\phi_i(\mathbf{y}) - \phi_i(G(\mathbf{x}))\|_1 \\
\mathcal{L}_{\text{style}} &= \mathbb{E}_{\mathbf{x},\mathbf{y}} \sum_{j} \| \text{Gram}(\psi_j(\mathbf{y})) - \text{Gram}(\psi_j(G(\mathbf{x}))) \|_1 \\
\mathcal{L}_{\text{stain}} &= \sum_{s \in \{H, E\}} \mathbb{E}_{\mathbf{y}} \| \text{Hist}(D_s(\mathbf{y})) - \text{Hist}(D_s(G(\mathbf{x}))) \|_1
\end{align*}

\subsection{Training Configuration}

All models were trained for a maximum of 50 epochs using the AdamW optimizer \cite{loshchilov2019decoupled} with learning rate $2 \times 10^{-4}$, $\beta_1 = 0.9$, $\beta_2 = 0.999$, and batch size 16. We employed cosine annealing learning rate scheduling with early stopping (patience=15) based on validation L1 loss. For VQGANs, codebook sizes were set to 1024 for each quantization level, with commitment costs of 0.2 for single-level and [0.2, 0.1] for the two-level hierarchy. The adversarial loss was introduced after 6000 iterations for the Orion dataset and 300 iterations for the smaller CODEX dataset, with weight 0.02 to ensure stable initial training. Gradient clipping (max norm 1.0) was applied to prevent training instability.

To ensure fair comparison across all models, we used an identical data preprocessing pipeline, including log transformation, outlier clipping (0.1-99.9 percentiles), and channel-wise z-score normalization for the mIF input. The best model was selected based on validation losses for all analysis in this work.

\subsection{Evaluation Metrics}
The crucial test for our models is their utility in real-world scientific analysis. We measured this by evaluating performance on two key downstream tasks. For nuclei segmentation, we assessed quality using the Mean Intersection-over-Union (IoU) \cite{everingham2010pascal} and the Dice Similarity Coefficient (DSC) \cite{dice1945measures} to measure the spatial overlap between predicted and ground-truth cell masks. For tissue classification consistency, we evaluated whether a pre-trained CONCH foundation model assigned the same tissue labels to generated images as it would to the corresponding ground truth H\&E images, measuring the semantic preservation of the virtual staining process. We also used a watershed algorithm \cite{vincent1991watersheds} for segmentation analysis.

%% file: texes/5_results.tex
\section{Results and Analysis}
\label{sec:results}

To rigorously evaluate our proposed architecture, we conducted a series of experiments on both the Orion and CODEX datasets, comparing our multi-level VQGAN against a single-level VQGAN and a standard conditional GAN (cGAN) baseline. We assessed performance through both image reconstruction metrics and, critically, on downstream tasks that measure the functional utility of the generated images for scientific analysis.

\subsection{Qualitative and Quantitative Image Reconstruction}

\begin{figure}[t!]
    \centering
    \includegraphics[width=\columnwidth]{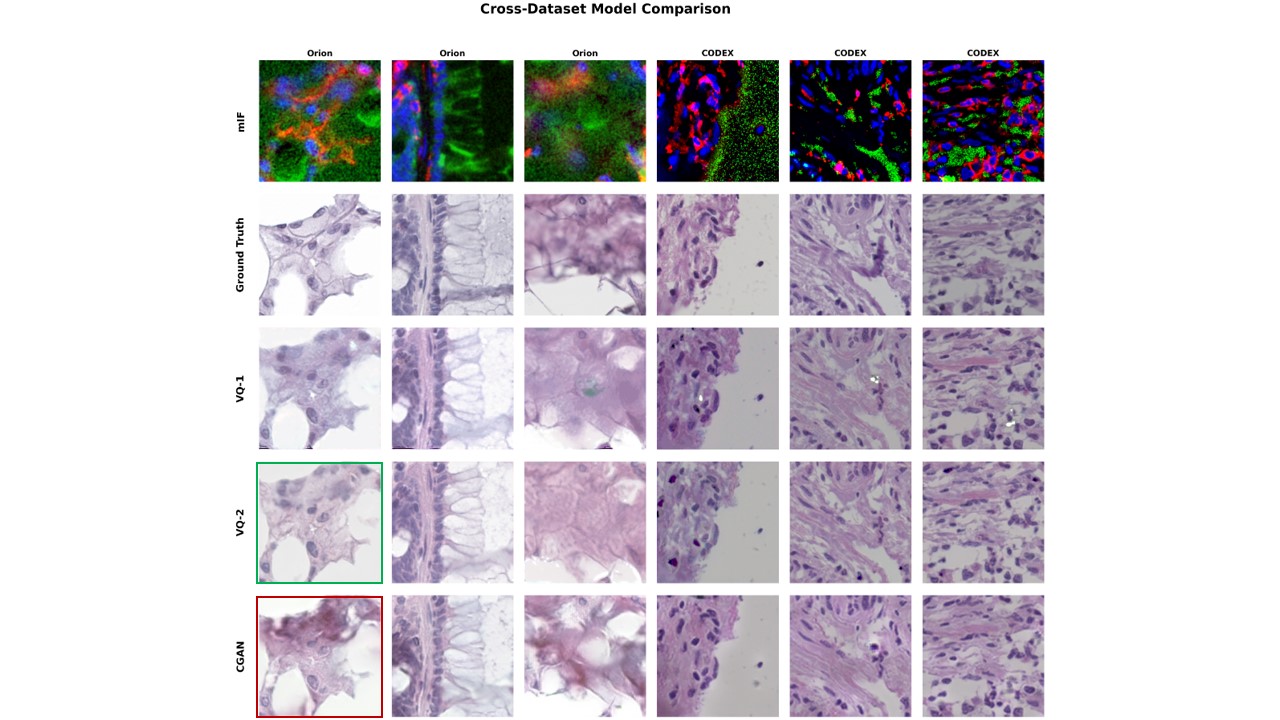} 
    \caption{\textbf{Qualitative Comparison of Virtual Staining Models on the Orion Dataset.} Each triplet shows the input mIF (red: CD31/CD45; green: E-cadherin/$\beta$-catenin; blue: Hoechst for Orion/CODEX respectively), the ground truth H\&E, and the generated H\&E from our proposed 2-level VQGAN. Our model successfully captures diverse morphological features, from glandular structures to dense cellular regions. We also highlight an example where the cGAN output shows degraded detail (marked in red), whereas our 2-level VQGAN preserves the structure better }
    \label{fig:qualitative_comparison}
\end{figure}

All three models were capable of generating visually plausible H\&E images that preserved the overall tissue architecture present in the input mIF data. A qualitative comparison of representative outputs is shown in Figure~\ref{fig:qualitative_comparison}. While all models produce high-quality translations, our proposed 2-level VQGAN consistently generates images with high structural fidelity and fewer color artifacts compared to the cGAN baseline.

Quantitatively, our 2-level VQGAN demonstrated superior performance across all standard reconstruction metrics on both datasets (Tables~\ref{tab:recon_metrics_orion} and \ref{tab:recon_metrics_codex}). On the Orion dataset, it achieved the lowest L1 and L2 reconstruction errors (0.1491 and 0.0391, respectively) and the highest Structural Similarity Index (SSIM) and Peak Signal-to-Noise Ratio (PSNR). On the CODEX dataset, the 2-level VQGAN similarly outperformed other models with the lowest reconstruction errors and highest similarity metrics. The hierarchical model showed consistent improvement over the 1-level VQGAN across both datasets, validating the benefit of its more expressive architecture for this task.

\begin{table}[h!]
\centering
\caption{Image reconstruction metrics on the Orion dataset. Best results are in \textbf{bold}.}
\label{tab:recon_metrics_orion}
\begin{tabular}{lcccc}
\toprule
\textbf{Model} & \textbf{L1} ($\downarrow$) & \textbf{L2} ($\downarrow$) & \textbf{SSIM} ($\uparrow$) & \textbf{PSNR} ($\uparrow$) \\
\midrule
cGAN & 0.1670 & 0.0495 & 0.4749 & 19.38 \\
VQGAN (1-level) & 0.1545 & 0.0435 & 0.5199 & 19.92 \\
\textbf{VQGAN (2-levels)} & \textbf{0.1491} & \textbf{0.0391} & \textbf{0.5221} & \textbf{20.38} \\
\bottomrule
\end{tabular}
\end{table}

\begin{table}[h!]
\centering
\caption{Image reconstruction metrics on the CODEX dataset. Best results are in \textbf{bold}.}
\label{tab:recon_metrics_codex}
\begin{tabular}{lcccc}
\toprule
\textbf{Model} & \textbf{L1} ($\downarrow$) & \textbf{L2} ($\downarrow$) & \textbf{SSIM} ($\uparrow$) & \textbf{PSNR} ($\uparrow$) \\
\midrule
cGAN & 0.1491 & 0.0395 & 0.6499 & 20.76 \\
VQGAN (1-level) & 0.1510 & 0.0449 & 0.6609 & 20.64 \\
\textbf{VQGAN (2-levels)} & \textbf{0.1432} & \textbf{0.0417} & \textbf{0.6816} & \textbf{20.89} \\
\bottomrule
\end{tabular}
\end{table}

\subsection{Downstream Functional Evaluation}

While reconstruction metrics are informative, a crucial test for virtual staining is whether the generated images are functionally useful for real-world scientific tasks. We evaluated this through downstream tissue classification and nuclei segmentation on both datasets.

\subsubsection{Tissue Classification}

To assess whether the generated images retain semantically meaningful features, we evaluated how well a pre-trained CONCH foundation model could assign the same tissue labels to generated images as it would to the corresponding ground truth H\&E images. This measures the semantic preservation of the virtual staining process.

\begin{figure}[h!]
    \centering
    \includegraphics[width=\columnwidth]{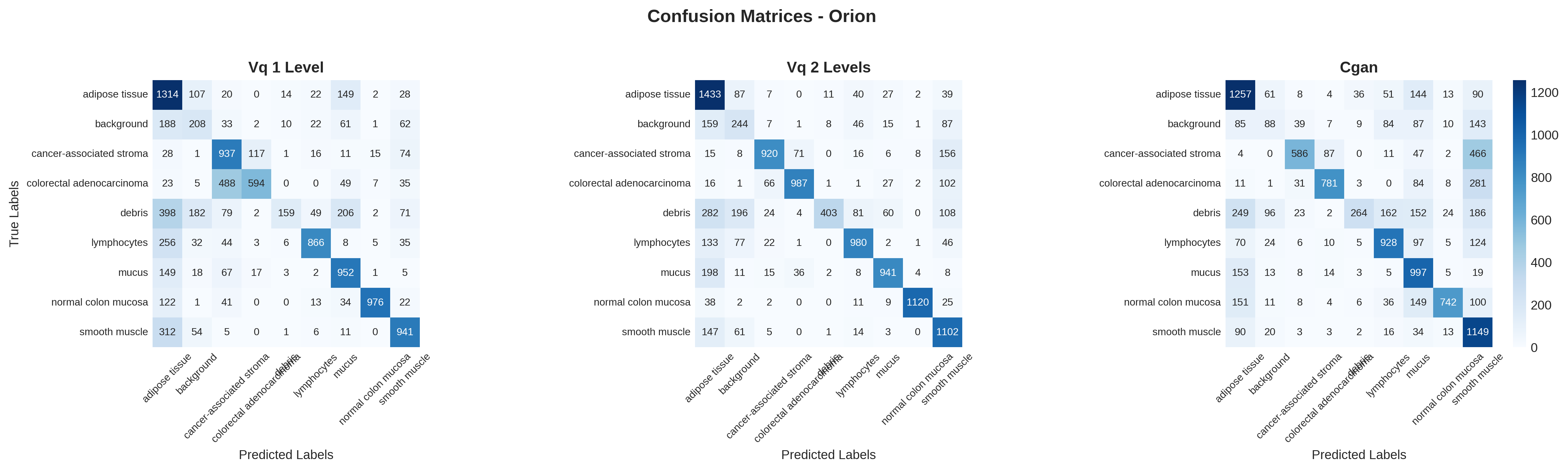} 
    \caption{\textbf{Tissue Classification Confusion Matrices.} The classifier performance on images generated by each respective model. Note the stronger diagonal for the VQGAN models, particularly the 2-level variant, indicating higher label consistency.}
    \label{fig:confusion_matrices}
\end{figure}

On the Orion dataset, the images generated by our VQGAN models achieved significantly higher label consistency than those from the cGAN. The 2-level VQGAN achieved the highest label agreement at 69.9\%, outperforming the cGAN baseline by over 11\%. On the CODEX dataset, the 1-level VQGAN achieved the highest label consistency at 53.5\%, outperforming both the 2-level VQGAN (48.4\%) and the cGAN baseline (39.9\%). The results are summarized in Table~\ref{tab:classification}. The confusion matrices in Figure~\ref{fig:confusion_matrices} further illustrate that the VQGAN models show a stronger diagonal and less confusion between distinct tissue types.

While these results validate the semantic preservation capability of our generated images, the performance gap from perfect agreement may reflect not only image quality but also the inherent sensitivity of foundation models to subtle domain shifts. The relatively modest label consistency rates may be attributed to an adversarial attack effect \cite{thota2024demonstration}, where small variations in the input images can significantly impact the performance of foundation models, highlighting their sensitivity to synthetic data distributions.

\begin{table}[h!]
\centering
\caption{Label consistency between generated images and ground truth H\&E using a pre-trained CONCH model. Best results for each dataset are in \textbf{bold}.}
\label{tab:classification}
\begin{tabular}{lcc}
\toprule
\textbf{Model} & \textbf{Orion} ($\uparrow$) & \textbf{CODEX} ($\uparrow$) \\
\midrule
cGAN & 0.6293 & 0.3987 \\
VQGAN (1-level) & 0.6742 & \textbf{0.5350} \\
\textbf{VQGAN (2-levels)} & \textbf{0.6994} & 0.4839 \\
\bottomrule
\end{tabular}
\end{table}

\subsubsection{Nuclei Segmentation}

We further evaluated functional quality by performing nuclei segmentation on both datasets. Interestingly, we observed that state-of-the-art deep learning-based segmentation models such as Cellpose \cite{stringer2021cellpose} and StarDist \cite{stevens2022stardist} showed poor agreement between the ground truth H\&E and generated images across all generative models, while traditional watershed segmentation maintained reasonable correspondence. This suggests that while the generated images are perceptually convincing, they may lack the precise, high-frequency textural cues that these advanced deep learning segmenters rely on, whereas traditional methods like watershed are more robust to such variations.

To enable a comparison, we instead used a traditional watershed algorithm \cite{vincent1991watersheds}, which is more robust to textural variations. On the Orion dataset, the 2-level VQGAN achieved the highest Mean IoU of 0.6800, outperforming both the 1-level VQGAN and the cGAN. This indicates that the hierarchical features learned by the 2-level model generate more distinct and separable cellular structures. The CODEX dataset showed that the 1-level VQGAN achieved the highest Mean IoU of 0.8192, outperforming both the 2-level VQGAN (0.8028) and the cGAN (0.7616). The complete results are shown in Table~\ref{tab:segmentation}.

\begin{table}[h!]
\centering
\caption{Downstream nuclei segmentation performance using a watershed algorithm. Best results for each dataset are in \textbf{bold}.}
\label{tab:segmentation}
\begin{tabular}{lcc}
\toprule
\textbf{Model} & \textbf{Orion Mean IoU} ($\uparrow$) & \textbf{CODEX Mean IoU} ($\uparrow$) \\
\midrule
cGAN & 0.6376 & 0.7616 \\
VQGAN (1-level) & 0.6270 & \textbf{0.8192} \\
\textbf{VQGAN (2-levels)} & \textbf{0.6800} & 0.8028 \\
\bottomrule
\end{tabular}
\end{table}

\begin{figure}[h!]
    \centering
    \includegraphics[width=0.6\columnwidth]{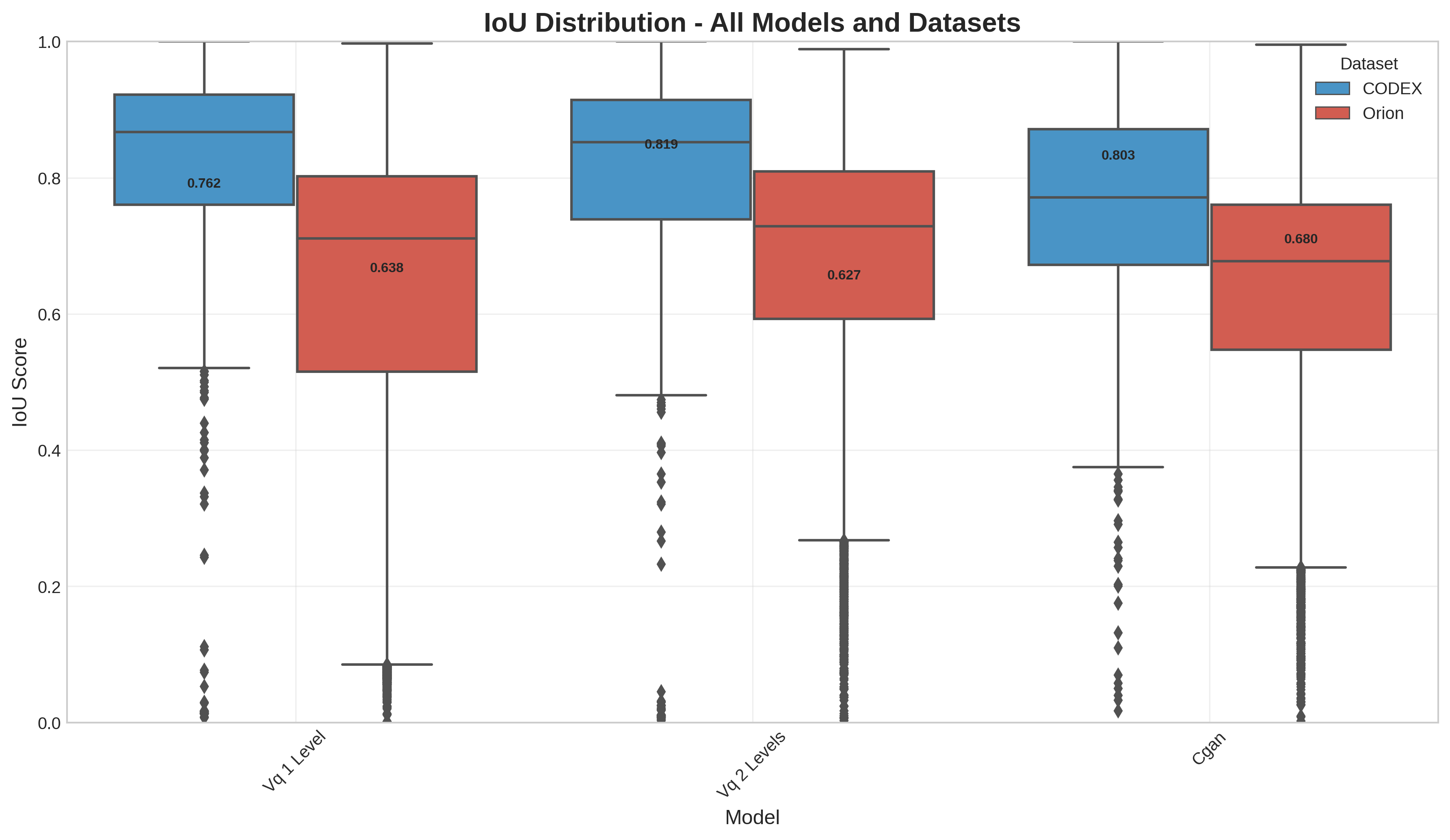} 
    \caption{\textbf{IoU Score Distribution for Nuclei Segmentation.} Box plots showing the distribution of IoU scores for each model on the Orion dataset. The 2-level VQGAN achieves a higher median score and a more favorable distribution.}
    \label{fig:segmentation_boxplot}
\end{figure}

\subsection{Ablation Studies}

We conducted several ablation experiments to optimize our VQGAN architecture and training procedures. We evaluated hierarchical quantization with 2, 3, and 4 levels of quantizers. Based on our experiments, we found that the 2-level architecture provided the best balance of performance and training stability, as increasing the number of codebook levels beyond 2 risked codebook collapse issues.

We also experimented with Macenko normalization \cite{macenko2009method} as an alternative preprocessing approach for the mIF input data. However, our standard channel-wise z-score normalization approach was retained for the final models based on the experimental results.

%% file: texes/6_conclusion.tex
\section{Conclusion}
In this work, we addressed the significant challenge of translating high-dimensional multiplex immunofluorescence (mIF) images into the standard morphological context of Hematoxylin and Eosin (H\&E) staining. We conducted a systematic comparison between a standard conditional GAN (cGAN) and a multi-level Vector-Quantized Generative Adversarial Network (VQGAN), evaluating their ability to produce high-fidelity and scientifically useful virtual stains.

Our results demonstrate that both architectural paradigms are capable of generating high-quality H\&E images. A key finding of this study is that a VQGAN, despite being constrained by a finite, discrete codebook, can successfully perform this complex translation task. This is a significant result, as it proves that a discrete latent space can effectively capture and compress the rich, high-dimensional information from multi-channel mIF inputs. We further validated the scientific utility of our generated images through downstream analysis, demonstrating that they are functionally robust for automated tasks like nuclei segmentation.

The success of the VQGAN framework suggests that its learned discrete codes could serve as a powerful foundation for other downstream tasks involving mIF data, such as quantitative cell classification or tumor microenvironment clustering. From a practical perspective, this research highlights the potential of virtual staining to rapidly generate an familiar H\&E view from complex molecular data, providing near-instant morphological context while waiting for time-consuming histopathology lab work. However, we emphasize that while these results are promising, the technology is still far from ready for direct clinical use. Extensive validation is required to ensure the diagnostic reliability and safety of AI-generated images in a clinical setting. Nevertheless, our study establishes that a multi-level VQGAN is a powerful and viable architecture for virtual staining, opening new avenues for integrating spatial biology with computational pathology.
